# Stub Wireless Multi-hop Networks using Self-configurable Wi-Fi Basic Service Set Cascading

Pedro Júlio', Filipe Ribeiro*, Jaime Dias*, Jorge Mamede', Rui Campos*
'INESC TEC and Instituto Superior de Engenharia do Porto – Instituto Politécnico do Porto, ISEP-IPP Porto
*INESC TEC and Faculdade de Engenharia, Universidade do Porto, FEUP
Porto, Portugal
{psjulio, filipe.a.ribeiro, jdias, jorge.mamede, rcampos}@inesctec.pt

*Abstract* — The increasing trend in wireless Internet access has been boosted by IEEE 802.11. However, the application scenarios are still limited by its short radio range. Stub Wireless Multi-hop Networks (WMNs) are a robust, flexible, and cost-effective solution to the problem. Yet, typically, they are formed by single radio mesh nodes and suffer from hidden node, unfairness, and scalability problems.

We propose a simple multi-radio, multi-channel WMN solution, named Wi-Fi network Infrastructure eXtension – Dual-Radio (WiFIX-DR), to overcome these problems. WiFIX-DR reuses IEEE 802.11 built-in mechanisms and beacons to form a Stub WMN as a set of self-configurable interconnected Basic Service Sets (BSSs). Experimental results show the improved scalability enabled by the proposed solution when compared to single-radio WMNs.

## I. Introduction

Wireless technology is currently a fast growing market driven by the demand of mobile Internet access. IEEE 802.11 based networks are now ubiquitous providing not only Internet access to mobile users, but also last mile network services in areas where wired networks are hardly an option, such as in rural, disaster, and temporary deployment scenarios. In these scenarios, IEEE 802.11-based Stub Wireless Multi-hop Networks (WMNs) are seen as a solution to enable robust, long range, and high bandwidth communications [1].

A Stub WMN is a static multi-hop wireless network composed of multiple Mesh Access Points (MAPs) placed between mesh clients and the wired infrastructure, as illustrated in Fig. 1. It consists of a static ad-hoc network with most of the traffic directed to and from the wired infrastructure. Each MAP operates both as transmitter, receiver, and relay forwarding packets to/from clients from/to other MAPs, and ultimately to the mesh gateway (GW). The main function of these devices is to provide Internet access to mesh clients (cf. Fig. 1). It is also possible to support other type of networks, such as sensor networks, cellular networks, and personal area networks. An example of such scenario is depicted in Fig. 1. Mesh connectivity is dynamically maintained along the network operation, as the Stub WMN is capable of self-configuring and self-healing around broken or congested links.

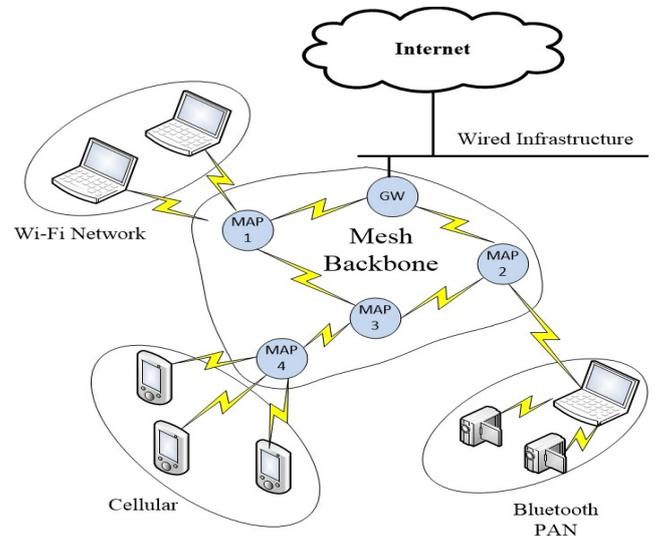

Fig. 1. Stub WMN reference architecture.

IEEE 802.11-based WMNs rely on the Carrier Sense Multiple Access with Collision Avoidance (CSMA/CA) mechanism for shared medium access control. CSMA/CA was designed having in mind single-hop networks. When used in multi-hop networks it leads to poor network performance – low throughput, high latency, and unfairness –, namely due to the hidden node problem; the use of RTS/CTS in multi-hop scenarios does not solve the problem and brings up the exposed node problem [1]. In order to work around this problem, multiple channels can be used. However, the channel switching process is time consuming, since it requires the local oscillator to be repositioned to each new frequency, resulting in increased latency and inefficiency. A simple way to overcome these problems is to equip nodes with multiple network interface cards (NICs). Yet, multi-radio operation is not straightforward and multiple factors have to be taken into account. The assumption that non-overlapping channels are non-interfering is not valid for mesh nodes equipped with multiple NICs, since *cross-talk* interfering effect causes significant performance drop on simultaneously active NICs operating on distinct frequency bands, when the NICs work with a physical separation lower than 46 cm [2] [3]. Also, there is a circular dependency between routing and Channel Assignment (CA), since one is heavily influenced by the other. For the best routes to be chosen the CA must be known, since it influences link capacity and network topology. However, CA must be aware of the expected load on each link, which is partially determined by the routing. In order to break this circular dependency, proposals have been made in the literature, either focusing on jointly solving the CA and routing problem or addressing them separately. Still, they rely on specific control messages, which leads to increased overhead and complexity; modifications to the MAC protocol have also been proposed [4], but they cannot be directly applied to commodity hardware.

We propose a novel distributed Joint Channel Assignment and Routing (JCAR) scheme, named Wi-Fi network Infrastructure eXtension – Dual Radio (WiFIX-DR). WiFIX-DR extends the WiFIX routing solution proposed in [1] to Multi-Radio Multi-Channel (MRMC) configurations. It reuses the IEEE 802.11 built-in mechanisms to create a tree-based WMN topology rooted at GW. The Stub WMN is formed by a set of interconnected IEEE 802.11 Basic Service Sets (BSSs) operating in different frequency channels. Each node acts as IEEE 802.11 Access Point (AP) for its children and as a Station (STA) of its parent. With WiFIX-DR there is no specific signaling messages issued by the nodes and medium access control is performed using CSMA/CA at each BSS only, avoiding the hidden node and exposed node problems faced by single-radio Stub WMNs.

Our major contributions are: 1) a **JCAR scheme for MRMC Stub WMNs built upon IEEE 802.11 mechanisms**, avoiding the overhead and complexity inherent to the state of the art JCAR solutions, which employ specific mechanisms and signaling messages; 2) a **mechanism to convey control information embedded in IEEE 802.11 beacons**, which is used by the MAPs to create the tree-based Stub WMN topology without any additional signaling messages.

The rest of the paper is organized as follows. Section II describes the state of art concerning MRMC solutions, presenting schemes focused on CA, routing, and joint CA and routing. Section III presents an overview of the WiFIX routing solution. WiFIX-DR is described in Section IV. WiFIX-DR implementation is described in Section V where we present the methodology used to insert information elements into beacons and disseminate control information throughout the Stub WMN. The assembled testbed used to evaluate its performance is presented in Section VI, along with results obtained concerning a set of performance metrics. Finally, Section VII draws the main conclusions and points out the future work.

## II. RELATED WORK

Prior work on MRMC WMN can be mostly divided into three major categories according to the problem they address: JCAR, Routing and CA. Given the interdependency between routing and CA, the JCAR problem is known to be NP-complete. Hence, state of the art solutions usually address each problem separately or address them sequentially.

Several proposals such as [2], [5] [6], and [7] have been made to solve the CA problem, which is known to be NP-hard [7]. Their main focus is on reducing interference among neighboring links to improve the aggregated capacity. Specifically, [7], [5] and [2] use link load as an input metric for the CA algorithm. These centralized approaches require dedicated mechanisms to transport the metrics all the way to the GW, whether through an additional interface, a dedicated channel, or both. The use of dedicated control NICs results in resource waste, since the control interface will not be used to its full potential. The same assumption is valid for dedicated control channels, since the actual amount of data flowing through these channels will only be a fraction of the total capacity available. In addition, these approaches limit the spectrum available for data exchange purposes. Hybrid approaches have also been proposed, such as [8] that uses a centralized clustering algorithm with distributed CA. Clustering approaches may be a good option for large-scale networks, however, they bring significant configuration complexity.

Routing strategies aim to find less congested paths in order to deliver packets in the shortest time possible. [9] is an example. It performs routing at Layer 2.5 and uses a flooding mechanism to announce the presence of a given node, in order to estimate the transmission quality, a metric used to select the best routes. Yet, the overhead introduced by the flooding mechanism is not negligible and may have a negative impact, namely on congested links. [10] adds QoS support to the OLSR routing protocol in order to deliver real-time traffic. When real-time traffic is to be forwarded, a logical topology is created based upon the session QoS parameters and the available bandwidth inferred from the dissemination of HELLO and TC messages. The layer wherein routing is performed is also an important factor. Layer 3 solutions such as [10] are tied to specific IP protocols. [11] algorithm employs a load balance technique that adapts to the QoS demands imposed by the upper layers. Nevertheless, it relies on a significant number of control messages that may degrade performance as the network grows.

Joint schemes have also been developed, such as [3], [12], and [13]. [3] and [12] consider the use of a channel switching methodology to optimize performance either to find better routing paths or update channels, respectively. However, this is a costly methodology, since it implies additional and time-consuming operations such as clock synchronization. A traffic profiling technique has also been proposed in [13]; however, the effectiveness of this JCAR solution relies on deep prior knowledge of traffic trends.

## III. WiFIX OVERVIEW

WiFIX [1] is a simple tree-based routing solution for Stub WMNs. It solves the route auto-configuration problem by configuring an active tree topology rooted at GW. This is accomplished by combining a single-message signaling protocol with the reuse of IEEE 802.1D learning bridges for frame forwarding at Layer 2. In order to support multi-hop Layer-2 forwarding, WiFIX defines an encapsulation method, named Ethernet over 802.11 (Eo11). Eo11 enables the establishment of virtual links (Eo11 tunnels) between neighboring MAPs on top of the physical IEEE 802.11 shared link. The Active Topology Creation and Maintenance (ATCM) mechanism is used to create the virtual links; together they form the active tree topology rooted at GW.

ATCM works as follows. GW periodically sends a Topology Refresh (TR) message, which is forwarded by any other MAP, after changing a set of parameters – number of hops to GW, parent address, and original address of the frame. Each MAP selects a parent node in the tree rooted at GW. The TR message is used to both announce GW and notify a MAP that it has been selected as parent in the tree. IEEE 802.1D bridges are used for packet forwarding on top of the active tree topology; they see the virtual links as regular Ethernet links.

Further details on WiFIX can be found in [1].

## IV. WIFIX-DR

WiFIX-DR is built upon the WiFIX routing solution and extends it to MRMC configurations. WiFIX-DR reuses the IEEE 802.11 built-in mechanisms to create a tree-based Stub WMN. The Stub WMN is formed by a set of interconnected IEEE 802.11 Basic Service Sets (BSSs) operating in different frequency channels. Both 2.4 and 5 GHz frequency bands are used to limit collision domains to one hop and avoid the *cross-talk* interference effect in each MAP. Each node acts as an IEEE 802.11 AP for its children and as an STA of its parent, as illustrated in Fig. 2. IEEE 802.11 beacon frames are used to convey the required signaling to auto-configure the Stub WMN, instead of the explicit TR messages defined in WiFIX. Medium access control is performed using CSMA/CA at each BSS only. Channels are assigned to interfaces based on a weight reduction algorithm, which improves channel diversity by increasing weight on a channel already used by parent nodes.

In what follows, we detail the IEEE 802.11 beacon-based signaling, the active topology creation approach, and the routing approach.

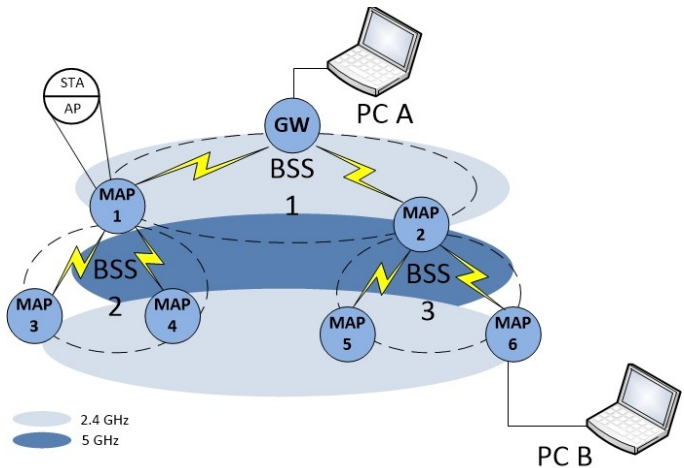

Fig. 2. WiFIX-DR architecture.

### A. Signaling Information Transport Mechanism

The IEEE 802.11 protocol already incorporates mechanisms to periodically broadcast information within a BSS. Specifically, APs use beacon frames to announce their presence and inform other STAs about which configurations should be used to connect to a BSS. WiFIX-DR leverages IEEE 802.11 beacons, with a technique called *beacon stuffing*, to broadcast metrics and enable BSS cascading that brings up the Stub WMN active topology.

One possible solution to embed information in beacons is to use the SSID field, which provides room for 32 bytes of information. However, this limits the application scenarios to those where the SSID field is not hidden. Another hypothesis is to use the BSSID field; however, it is only 6 bytes long. The 191 empty bits in length fields of the Information Elements (IE) have also been explored for information embedding [14]. Still, mapping variable network metrics into these fields may be a complex task and may not be feasible. The vendor specific IE is an IEEE 802.11 standard and allows each vendor to add specific information to beacon frames. Each of these elements is limited to 252 bytes of payload (256 bytes in total minus 3 Organizationally Unique Identifier (OUI) bytes minus 1 byte to specify the type). Additional vendor's IEs may be added as long as the beacon frame body does not exceed 2320 octets. Since overloading the vendor elements is the most flexible solution, we used it for signalling within a WiFIX-DR Stub WMN. This approach has also the advantage of enabling the modification of the information contained in the IE without breaking the associations between each AP and the attached STAs. The general format of the vendor specific IE is depicted in Fig. 3.

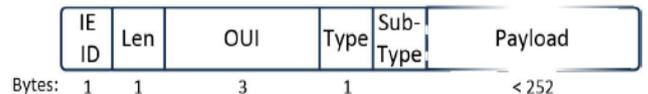

Fig. 3. Vendor Specific Information Element.

The ID 221 identifies this IE and it is always the last IE in an IEEE 802.11 beacon. The length field follows. As in other IEs it sets the size of the information carried. This includes OUI, type, sub-type, and payload. The OUI field contains the globally unique identifier assigned to each organization by the IEEE Registration Authority. We adopted an OUI that was not yet assigned to any organization, *FF-FE-00*. We set the type field to a constant value of *01*, while the subtype is overwritten with the topology information, namely the number of hops to the GW (HOPs in Fig. 4). The payload field contains the actual data carried by beacons. Here, we place an ordered list of channels used by the upstream neighbors, with each MAP adding to the end of the list the channel assigned to its *UP-NIC*.

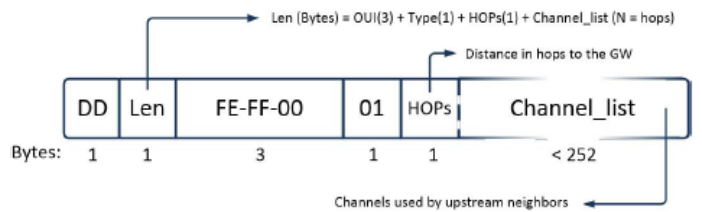

Fig. 4. WiFIX-DR modified Vendor Specific IE.

### B. Active Topology Creation

The hierarchical topology rooted at GW is created in two major steps: *1) Interface to neighbor binding*; *ii) Channel assignment*. GW is the exception to this process, since it is equipped with only one interface and makes a selfish CA based on the *Automatic Channel Selection (ACS)* algorithm, a survey based algorithm, used by default by *hostapd* based APs, that estimates the best channel on a given band based on occupancy and perceived noise floor.

***i) Interface to neighbor binding.*** MAPs with no directed path to the GW passively listen to beacons broadcast by nearby MAPs.

First, only the GW beacons are available with *HOPs=0* and *Channel list=NIL*. Once there is a valid set of candidate parents, the current MAP selects the parent that guarantees the shortest path to the GW. After selecting a parent, the MAP configures the corresponding NIC, named *UP-NIC*, in STA mode and triggers the IEEE 802.11 standard association with the parent. Next, the MAP configures the other NIC, named *DOWN-NIC*, in AP mode, so that a new BSS is created and associations from children MAPs can be accepted. The *UP-NIC* and *DOWN-NIC* are always configured in different frequency bands, for the reasons explained above. The overall process is repeated by each MAP in the Stub WMN.

*ii) Channel assignment.* Each MAP creates a list of candidate channels for configuring the *DOWN-NIC*. This list contains all non-overlapping channels of the frequency band not used by the parent (2.4 GHz or 5 GHz). A weight reduction algorithm is used to choose the best channel. Initially, the weight of each channel in the chosen band is set to 1. Then, the weight is updated for each channel $k$ found in *Channel list* of the beacons broadcast by the selected parent, according to Eq. 1.

$$weight_{k\_new} = weight_{k\_current} * (d_k/d_{hops}) \qquad (1)$$

where $weight_{k\_current}$ represents the current weight of a channel used at $d_k$ hops from the current MAP which is $d_{hops}$ apart from GW, where $d_k/d_{hops}$ is always lower than 1. At the end, the channel with the highest weight is assigned to the *DOWN-NIC*.

This simple mechanism aims to improve channel-diversified paths while ensuring that the equal frequency channels are assigned with the maximum hierarchical spatial separation.

*C. Routing*

WiFIX-DR routing scheme resorts to the mechanisms already provided by WiFIX to forward traffic over Layer-2. Virtual links are established based on Eo11 tunnels with an 802.1D learning bridge at each tunnel endpoint. The establishment of new virtual links is triggered upon the completion of the standard IEEE 802.11 association process of a node to its parent. The regular data exchange from a node in the mesh feeds the learning bridge algorithm. Hence, routes to a newly joined node in the mesh are discovered once it starts to generate traffic, as in any switched Ethernet Local Area Network (LAN).

V. IMPLEMENTATION

WiFIX-DR was implemented reusing most of the functionalities provided by the original WiFIX implementation [1]. The module runs in user-space and invokes most of the functionalities provided by the Linux Operating System (OS).

The *hostapd* and *wpa_supplicant* modules coordinate the AP and STA functionality of each interface. To enable the insertion of metrics into beacons a communication channel is established between WiFIX-DR and *hostapd* using the provided Linux socket file for control purposes. Kernel events are monitored with *netlink* sockets, which in turn establish or terminate the Eo11 tunnels. The establishment of tunnels resorts to Layer-2 virtual interfaces (taps) that connect to the 802.1D learning bridge, as it happens in WiFIX.

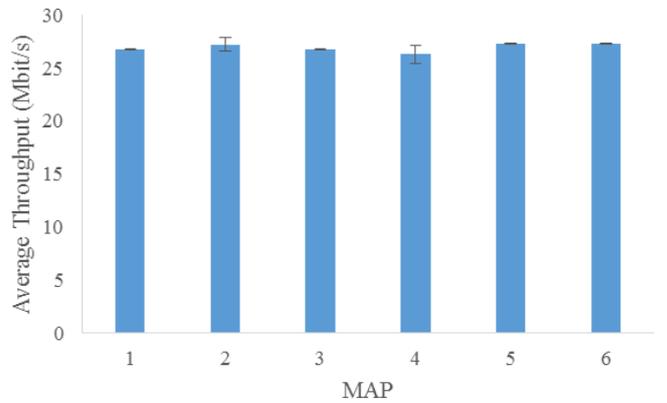

Fig. 5. Average UDP throughput between each MAP and the GW

One of the main enhancements when compared to the original WiFIX implementation is the absence of control traffic by removing the ATCM mechanism. The creation of taps in WiFIX resorts to ATCM mechanisms, notifying remote endpoints of new nodes at their parenthood. To make the tap list consistent and updated as new connections are established and terminated without any explicit message in WiFIX-DR, a new method had to be developed. The table of stations associated to an AP is accessible from user space, so one option could be to perform periodic polls and parsing its content for any modifications. However, polling has to be well balanced; the increase in polling periodicity augments the waste in processing time if no modification happened to the table between polls, and low periodicity leads to an increase of time to make the tunnel available to route traffic, increasing the probability of packet loss. The other option is to receive kernel events related to associations and disassociations. This is the approach used by WiFIX-DR. In this away the MAP list is updated almost simultaneously with the associated stations list avoiding all polling disadvantages. *Netlink* application programming interface provides the means to establish connections between user and kernel space. It also enables processes to multicast or listen to multicast groups of the *netlink* family it is connected to, with no need to implement additional features in kernel space.

By subscribing to the *nl80211* group of the *netlink generic* family, the algorithm will listen to the wireless system kernel events and filter them to select only the ones concerning MLME, since we want to get event notifications related to associations and disassociations. More specifically, from the list of all possible events two are of the most interest: the *NL80211_CMD_NEW_STATION* is issued, at a parent, when connections are established, so a new TAP is consequently added to the tap list with the MAC address of the newly connected stations. Conversely, when a station disconnects, the command *NL80211_CMD_DEL_STATION* is received from the kernel and the corresponding TAP removed from the list.

Upon the establishment of tunnels, data is able to flow between endpoints. Before being sent over the real interface, at each hop, data is first forwarded to the virtual learning bridge that will map the port of a tap (or multiple if destination is the broadcast address) as the next hop. According to the intended destination WiFIX-DR then encapsulates the frame, setting source to be the current node's MAC address and destination to correspond to the tap (next hop) address. The interface for data is selected according to the next hop towards the destination. Because at each MAP the two real interfaces are associated with different levels of the tree hierarchy, the selection follows a binary criterion, since the next hop is either the parent or any of the children. An exception occurs when, for example, the destination is unknown and broadcast ARP requests must be directed in both directions.

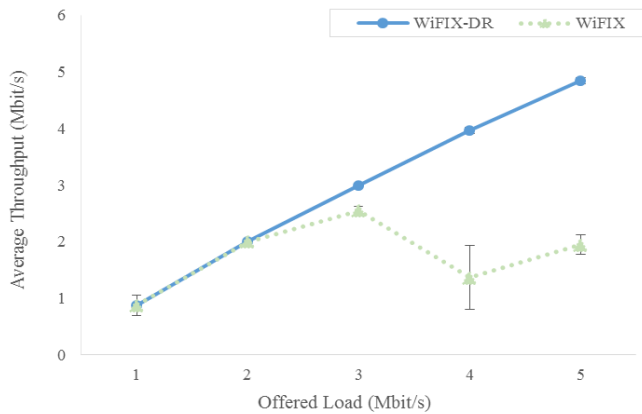

Fig. 6. Average one-way UDP throughput versus offered load per MAP.

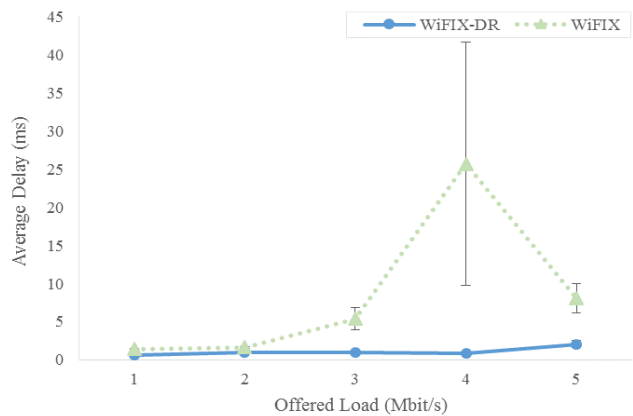

Fig. 7. Average one-way UDP delay versus offered load per MAP.

Frames received on the real interface are forwarded to WiFIX-DR via the packet socket to proceed to decapsulation. If the recipient is not the destination, the frame will be forwarded to the next hop with a new frame header to match the current link endpoints.

## VI. EXPERIMENTAL EVALUATION

The performance of WiFIX-DR was evaluated experimentally using the implementation described in Section V and a 7-node Stub WMN, as depicted in Fig. 2. The conducted tests aimed to evaluate WiFIX-DR against WiFIX by considering throughput and delay as performance metrics. The MAPs were implemented using *TP-Link Archer c5 v1.2* routers supporting two wireless cards, operating both at 2.4 GHz and 5GHz frequency bands, and running *OpenWRT 15.05 Chaos Calmer rc3*. The wireless NICs at each MAP run on IEEE 802.11g (2.4 GHz) and IEEE 802.11a (5 GHz) and used fixed data rates (54 Mbit/s). The IEEE 802.11n could also be used as basis, since WiFIX-DR is IEEE 802.11 variant agnostic. *Iperf3* was used to generate UDP flows from a laptop connected to each MAP to a laptop connected to the GW. For each MAP, ten tests were performed, each with a duration of 60 seconds. The packet size was set to 1400 bytes and tests began after having the full topology established. The IEEE 802.11 beacon frame interval was set to 100 ms. Nodes were deployed in an indoor scenario with the transmission power set to 1 mW with an average received signal strength of -60 dBm between a node and its parent.

We started by examining the maximum achievable throughput of individual GW paths. To perform this experiment only the path being tested was active at each moment. The offered load was set to saturate the IEEE 802.11g link. As depicted in Fig. 5, both nodes at one hop and two hops from GW achieve the same throughput; this is in fact the main advantage of the MRMC approach. By utilizing multiple non-overlapping channels at each hop we limit the collision domains to one hop. Hence, communications at each hop can run simultaneously without collisions and any contention.

The performance of WiFIX-DR solution was also studied with simultaneous flows and compared with the single-radio version using WiFIX. For this test case, one UDP flow was generated in each MAP towards GW with the bitrate increasing from 1 to 5 Mbit/s in 1 Mbit/s steps. The plot in Fig. 6 shows the average throughput comparison between WiFIX-DR and WiFIX. Both solutions present similar values for data rates below 2 Mbit/s. At this point WiFIX reaches saturation while WiFIX-DR keeps a steady growth with the average throughput matching the offered load until the saturation point is reached at 5 Mbit/s. As expected, the channel diversity provided by WiFIX-DR solution leads to the overall capacity improvement, due to the reduction in the number of competing flows and intra-path interference. Ultimately, only nodes at the same hierarchical level compete for a transmission opportunity.

Fig. 7 presents the delay results obtained for both solutions. They follow the same trend observed for the throughput. The lower delays achieved by WiFIX-DR are a direct consequence of the higher network capacity enabled by the limitation of collision domains to one hop.

## VII. DISCUSSION

The performed tests clearly show the improvement of WiFIX-DR when compared to WiFIX. MAPs joining the network are broadly aware of channels used by upstream neighbors and use this knowledge to join the network with minimum impact to the overall performance. By leveraging IEEE 802.11 beacon frames to convey signaling information, the medium is only occupied for data forwarding purposes and standard IEEE 802.11 signaling. Such improvement is achieved inexpensively by equipping nodes with two interfaces and taking advantage of full spectrum provided by the 2.4 GHz and 5GHz frequency bands. In its current version, WiFIX-DR is unaware of path performance degradation, which inhibits it to search for better routing paths. The CA algorithm's decision is made exclusively with information provided by upstream neighbors. Thus, interference may arise from BSSs at the same hierarchical level. Since lower levels tend to be denser, this will translate in higher performance degradation for MAPs located further away from GW. On the other hand, this approach reduces interference on BSSs, which are closer to GW and located in the hotspot zone of the Stub WMN. These limitations and tradeoffs shall be addressed in the next version. Nevertheless, the obtained results show that WiFIX-DR is able to improve scalability while maintaining performance independently from the node's distance from GW. Furthermore, WiFIX-DR ensures the same throughput fairness found in an IEEE 802.11 BSS.

## VIII. CONCLUSION

Wireless Internet access has been boosted by IEEE 802.11. However, IEEE 802.11 has limited radio range. Stub WMNs are a robust, flexible, and cost-effective solution to extend its coverage. Yet, typically they are formed by single radio MAPs, with the network capacity decreasing with the number of hops. Herein, we proposed WiFIX-DR, a simple multi-radio, multi-channel solution built upon WiFIX and IEEE 802.11 beaconing and AP-STA operation mode. With WiFIX-DR network capacity remains the same as the number of hops increases; this has been validated experimentally. Additionally, throughput fairness is ensured, as the use of CSMA/CA for medium access control is limited to each BSS forming the Stub WMN.

As future work, we will focus on the improvement of routing and CA strategies to address both of the constraints afore mentioned. We will also add the mechanisms to enable the network self-reorganization. Moreover, we will test WiFIX-DR in a larger and denser testbed and compare to other state of art solutions.